\documentclass[a4paper]{article}
\usepackage{indentfirst} 
\usepackage{bm,multirow}
\usepackage{graphicx} 
\usepackage{url}
\usepackage{fancyhdr}

\usepackage{geometry}
\begin{document}
\thispagestyle{empty} \vspace*{0.8cm}\hbox
to\textwidth{\vbox{\hfill\noindent \\ \textit{Proceedings of the 8th International Conference on Pedestrian and Evacuation Dynamics (PED2016)\\
Hefei, China - Oct 17 -- 21, 2016\\
Paper No. 12}
\hfill}}
\par\noindent\rule[3mm]{\textwidth}{0.2pt}\hspace*{-\textwidth}\noindent
\rule[2.5mm]{\textwidth}{0.2pt}

\begin{center}
\LARGE\bf Pedestrians’ rotation measurement in bidirectional streams 
\end{center}

\begin{center}
\rm Claudio Feliciani $^1$ \ and \ Katsuhiro Nishinari $^{2,3}$
\end{center}

\begin{center}
\begin{small} \sl
${}^{\rm 1}$ Department of Advanced Interdisciplinary Studies, Graduate School of Engineering, \\ The University of Tokyo, 4-6-1 Komaba, Meguro-ku, Tokyo 153-8904, Japan \\
feliciani@jamology.rcast.u-tokyo.ac.jp \\
${}^{\rm 2}$ Department of Aeronautics and Astronautics, Graduate School of Engineering, \\ The University of Tokyo, 7-3-1 Hongo, Bunkyo-ku, Tokyo 113-8656, Japan \\ 
${}^{\rm 3}$ Research Center for Advanced Science and Technology, \\ The University of Tokyo, 4-6-1 Komaba, Meguro-ku, Tokyo 153-8904, Japan \\
tknishi@mail.ecc.u-tokyo.ac.jp
\end{small}
\end{center}
\vspace*{2mm}

\begin{center}
\begin{minipage}{15.5cm}
\parindent 20pt\small
\noindent\textbf{Abstract - } This study presents an experimental measurement of pedestrians’ body rotation in bidirectional streams. A mock-up corridor monitored using a camera placed on azimuthal position is used to study pedestrians' behavior in unidirectional and bidirectional flows. Additionally, a commercial tablet is fixed on the chest of sample pedestrians to examine their body rotation (or yawing) which cannot be obtained using position tracking alone. Angular velocity is recorded and simultaneously stored in a central location using a wireless network, thus allowing the analysis of body movements with a high sampling rate and a limited delay. To investigate the influence of major/minor flow proportion (flow-ratio) on bidirectional streams two different situations were tested: the balanced configuration (with equal flows in both directions) and an unbalanced configuration (with different major and minor flow). Results clearly show that unidirectional flow is more stable compared to the bidirectional case, requiring less time to cross the experimental section and showing a very small amount of rotation during the whole experiment. Both bidirectional configurations showed high values of body rotation, in particular during lane formation and dissolution. Finally, rotation directly measured on pedestrians' body was compared with the one obtained indirectly by analyzing pedestrians' trajectories. The comparison shows that, at least from a qualitative point of view, both methods are in agreement, thus suggesting that even properties which can only be measured by motion sensing could be obtained indirectly through the analysis of trajectories. Concluding, it has been suggested that while lanes help smooth out bidirectional flows, larger instabilities are observed compared to the unidirectional case. Lane separation and/or appropriate guidance are therefore required. In this regard, the development of a dynamically changing guidance system is a scientific and technical challenge on which further research could be addressed.
\end{minipage}
\end{center}

\begin{center}
\begin{minipage}{15.5cm}
\begin{minipage}[t]{2.3cm}{\bf Keywords:}\end{minipage}
\begin{minipage}[t]{13.1cm}
bidirectional flow, counter flow, body rotation, lane formation, tablet, gyroscope
\end{minipage}\par\vglue8pt
\end{minipage}
\end{center}

\section{Introduction}  
Pedestrian crowds are a common sight in large cities all around the globe and with the increasing urbanization an efficient system to manage them is becoming necessary. While predicting the motion of a large number of people in complex structures such as train stations or airports requires specialized algorithms considering the interactions between pedestrians and the surrounding environment, there are some common constructional features which can be considered separately with a sufficient degree of generalization. Corridors, corners and crosswalks are an ubiquitous feature in urban infrastructure constituting the connections between different areas inside a building or different parts of a road network. Corridors (or more in general straight sections) are found everywhere and comparisons can be performed in a fairly accurate way. \par
While research on pedestrian dynamics mostly started from bottlenecks and simple bidirectional streams \cite{Blue1999,Helbing1995,Weidmann1993}, the evolution of the simulation algorithms and the necessities of the industry have shifted the focus toward increasingly more complex phenomena. Nonetheless some of the basic principles of pedestrian dynamics are still mysterious and experimental research is required to understand the mechanisms leading to crowd accidents. For the particular case of the bidirectional flow, different authors have performed a number of studies which are briefly summarized as follows. \par
Lam et al. \cite{Lam2002} investigated crosswalks in Hong Kong and derived a function for the effective capacity in function of the flow ratio. They concluded that because of the formation of lanes, balanced bidirectional flow (same amount of flow in each direction) is equivalent to two unidirectional flows moving in opposite directions. When the flow in both directions is different, the instability of the minor flow contributes to the reduction of the capacity on that direction. Alhajyaseen et al. \cite{Alhajyaseen2011} performed a similar analysis but they were able to additionally distinguish between different age groups. \par
Kretz et al. \cite{Kretz2008} collected valuable experimental data for bidirectional flow in a corridor and they concluded that bidirectional flow has always an higher capacity compared to unidirectional flow. Kaufman \cite{Kaufman2007} also performed observations of bidirectional flow with a special attention paid on the number of lanes formed. He concluded that the number of lanes depends on the overall stability of the flow and that short-term counter-flow situations leads to a number of lanes larger than 2. Nowak et al. \cite{Nowak2012} studied the stability of bidirectional flow using a cellular automaton model in which they distinguished between different flow states: free-flow, disorder, lanes and gridlock. The same type of conclusions were experimentally obtained by Feliciani et al. \cite{Feliciani2015} in an empirical observation. Zhang et al. \cite{Zhang2012} performed one of the largest supervised experiments for bidirectional flow involving up to 350 people. They concluded that while differences between uni- and bidirectional flows can be found examining the fundamental diagrams, the differences for different bidirectional flow configurations are small. \par
In our previous research we focused on the behavior of pedestrians in bidirectional flow using both simulation \cite{Feliciani2016} and supervised experiments \cite{Feliciani2016pre}. By performing a number of experiments with different configurations we found 5 different phases in bidirectional flows and our analysis showed that body rotation has an important role in the process of lane formation. However, our analysis was completely based on the tracking obtained from head position, which cannot deliver accurate information on the amount of body rotation. \par
A few studies in the past \cite{Roggen2011,Bujari2012,Kjaergaard2012} already considered the use of smartphones or mobile sensors to gain information from the crowd. However, in most of the cases only the accelerometer was used to collect data and the main scope has been detection of flock formation or recognizing of common patterns among pedestrians. \par
In the study presented here we measure body rotation directly on pedestrians by making use of the inertial measuring unit (IMU) contained in commercial tablets which were fixed on pedestrians' chest. In the following chapters we will introduce the technical details for such a measurement and discuss the main findings obtained from data analysis.

\section{Experiment design and procedure}
To study the motion of pedestrians in bidirectional flow a walking path delimited by band partitions was created on a sidewalk in an outdoor area closed to traffic (see Fig.1 and Fig.2). The width of the section was 2.4 m and the area considered for the observation (measurement area) had a length of 10 m. 

\vspace*{4mm} 
\centerline{\includegraphics[angle=90,trim=8cm 3cm 6cm 3cm,clip,width=14cm]{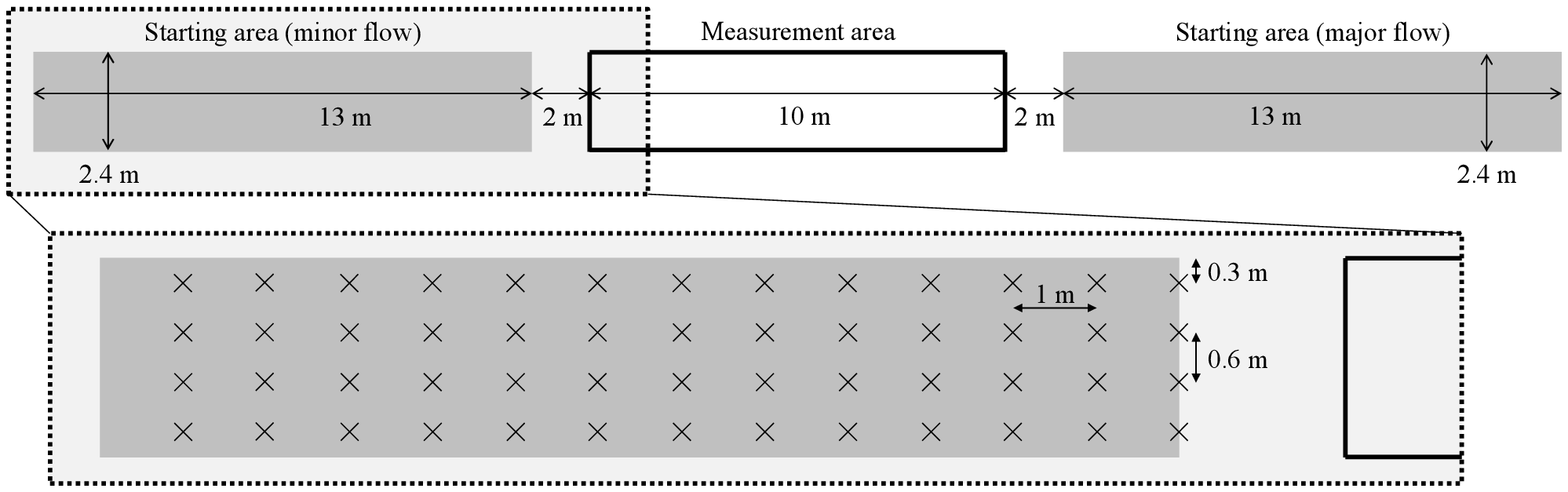}}
\begin{center}
\parbox{15.5cm}{\small\center{\bf Fig.1.} Geometrical setup of the experiment.}
\end{center}
\vspace*{4mm}

Two starting areas were created in which pedestrians had to take specific positions before the start of each experiment. Between the central measurement area and each of the two starting areas a distance of 2 m was set allowing pedestrians reaching a constant walking speed before entering the central part. In each of the two starting areas 52 crosses were drawn on the ground at uniform distance. In the first column (at a distance of 2 m from the start of the central section) 4 positions were created at 60 cm spacing (30 cm from the external side of the section). Additional columns were created at a longitudinal distance of 1 m from each other's. \par
50 male university students took part to the experiment and received financial reward for the time spent. Each of the participants received a black T-shirt and caps of different colors (details will be given below). The experiment was performed on November 14th 2015 under cloudy, changing weather conditions with an average temperature of about 13$^{\circ}$C. Although some light rain fell before, during and after the experiment, the results reported here were taken under dry, cloudy weather conditions.  \par
A camera was placed on azimuthal position at an height of 21 m pointing at the center of the measurement area. Video recordings of the experiments were taken at full HD mode (1920 x 1080 pixel) with a frame rate of 30 fps. However, only the central part was considered for data analysis, thus reducing the size of the video analyzed to an area of about 885 x 220 pixel. An example for a frame taken from the experiment is given in Fig.2. Pedestrian recognition and tracking were performed based on the color of each participant's hat by using PeTrack software (version 0.8) \cite{Boltes2013}.

\vspace*{4mm} 
\centerline{\includegraphics[angle=0,trim=0cm 0cm 0cm 0cm,clip,width=9.5cm]{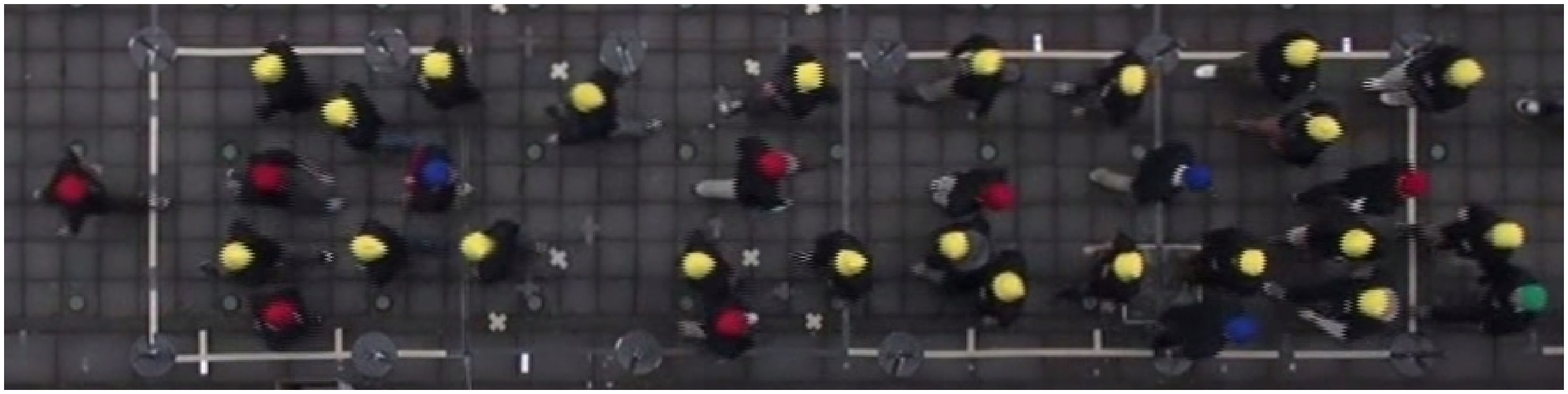}}
\begin{center}
\parbox{15.5cm}{\small\center{\bf Fig.2.} Example for a video frame taken from the unbalanced bidirectional flow experiment. }
\end{center}
\vspace*{4mm}

Different flow configurations were tested during the one hour experiment. At first a unidirectional configuration was tested by having the participants taking all the positions available in the right starting area (two spots were unused in the last column as the total number of positions exceed the number of participants by 2). Fig.3 shows a schematic example for the starting positions used during the unidirectional configuration. Participants belonging to the major flow wore yellow caps (light gray in Fig.3) and those equipped with a tablet to measure their movements had green caps (dark gray in Fig.3). The unidirectional configuration was tested 4 times.

\vspace*{4mm} 
\centerline{\includegraphics[angle=90,trim=7.7cm 0.5cm 7.5cm 0.5cm,clip,width=10cm]{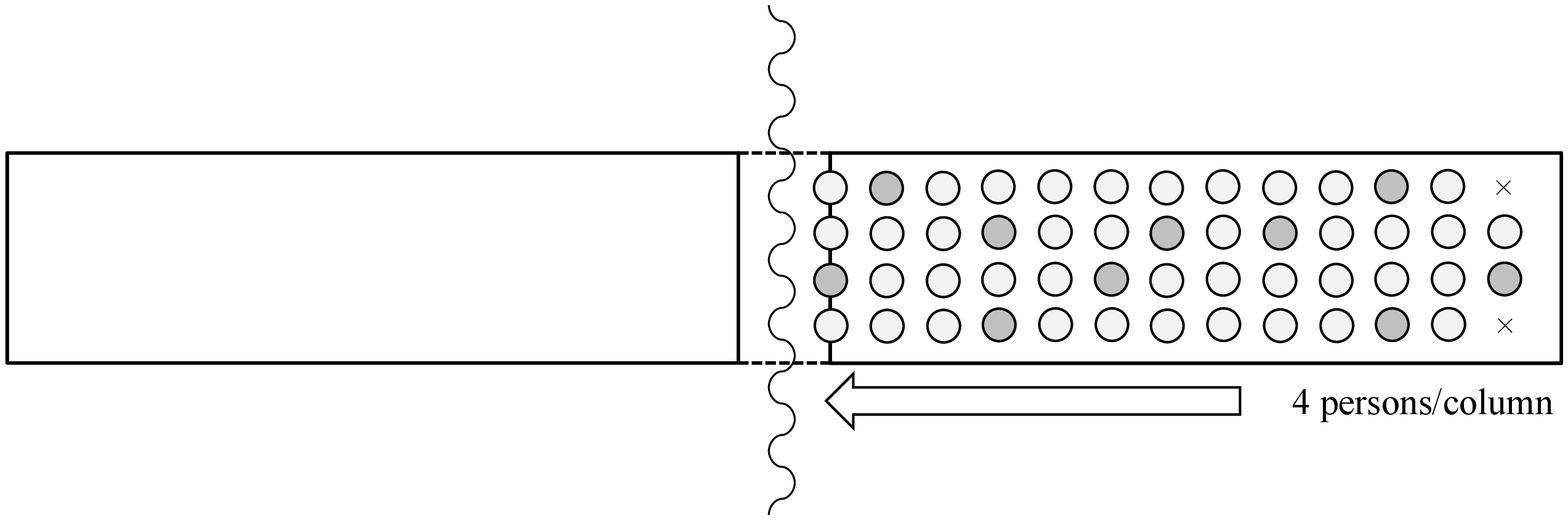}}
\begin{center}
\parbox{15.5cm}{\small\center{\bf Fig.3.} Unidirectional configuration. }
\end{center}
\vspace*{4mm}

It is important to remark that starting position examples given here are only intended for explicative purpose because actual positions taken by participants changed from one run to the other and participants were specifically asked to avoid taking always the same starting position. In addition, staff was instructed to reshuffle participants and to check for the ones staying close to friends/colleagues and/or staying always on similar positions. Conditions were specifically created to avoid participants getting used to the experiment being conducted and after analyzing the videos we could conclude that a learning process did not take place (the instability of the weather and the few interruptions caused by rain probably contributed in avoiding quick learning of the experimental procedure). \par
Next, an unbalanced bidirectional flow configuration was tested having a small portion of participants moving from the right starting area to the left side (see Fig.4 above). In the major flow (right) participants were allowed to take 3 positions for each column (one spot was left empty). On the minor flow side (left) participants were allowed to take one position for each column. The configuration was tested 4 times. Again, we asked the participants to change their position on each repetition and staff was present on each starting area to check for randomness of the starting layout. Simple pedestrians of the minor flow wore red caps (represented as dark gray dots in Fig.4), while pedestrians equipped with tablets had blue caps (black dots in Fig.4). \par
Finally a balanced bidirectional flow configuration was tested by having the participants taking 2 positions for each column of the starting areas. Needless to say, starting position were changed during each run and the experiment was repeated 5 times, although only 3 could be considered as valid because some participants stopped to adjust band partitions alignment during 2 executions. The 3 different configurations are summarized in Tab.1. 

\vspace*{4mm} 
\centerline{\includegraphics[angle=90,trim=7.7cm 0.5cm 7.5cm 0.5cm,clip,width=10cm]{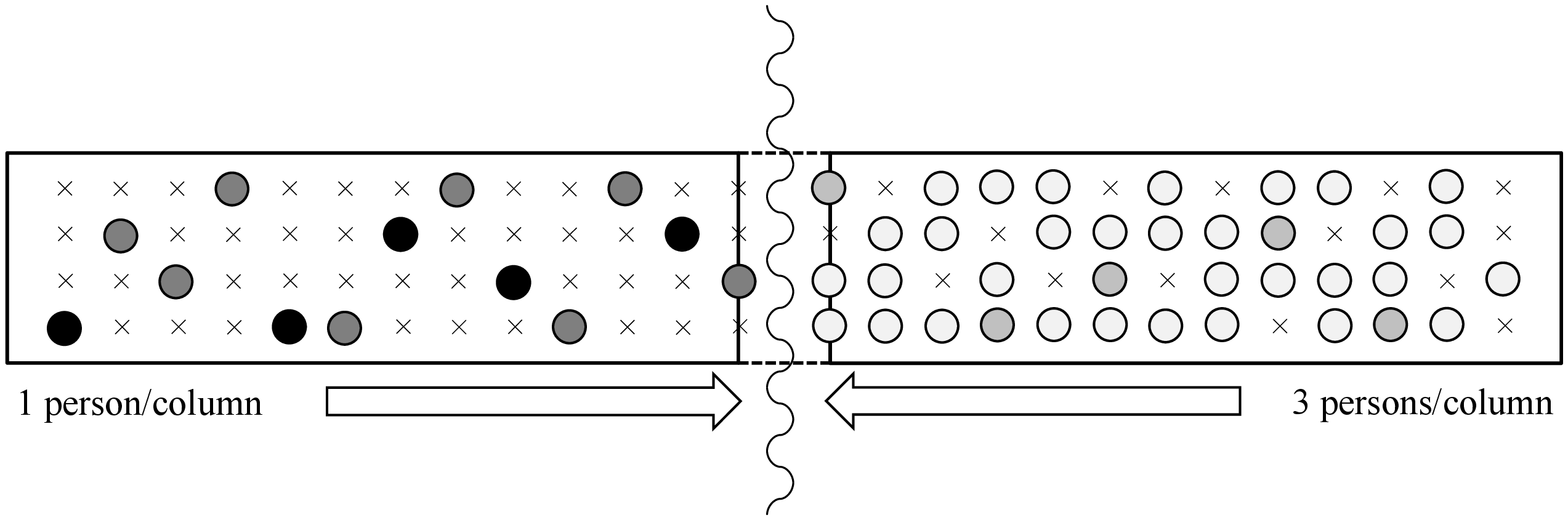}}
\vspace*{2mm}
\centerline{\includegraphics[angle=90,trim=7.7cm 0.5cm 7.5cm 0.5cm,clip,width=10cm]{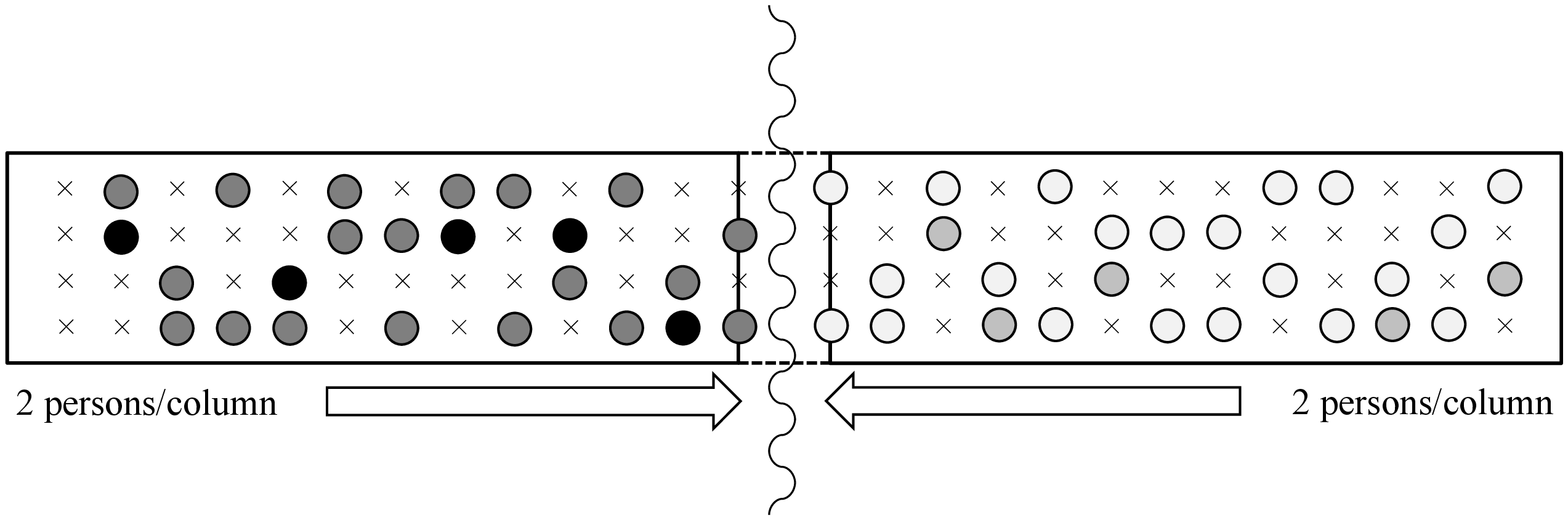}}
\begin{center}
\parbox{15.5cm}{\small\center{\bf Fig.4.} Unbalanced (above) and balanced (below) bidirectional flow configuration. }
\end{center}
\vspace*{4mm}

\begin{center}
Table 1: Experimental configurations and details
\end{center}
\begin{table}[htbp]
\begin{center}
\begin{tabular}{|l|l|l|l|l|}
\hline 
Configuration name  						& Code 				& Major flow 			& Minor flow 				& Valid runs		\\
\hline 
Unidirectional flow 						& 4/0 (UNI) 	& 4	pers./column	& --								& 4							\\
Unbalanced bidirectional flow 	& 3/1 (BI)		& 3 pers./column	& 1	pers./column		& 4							\\
Balanced bidirectional flow 		& 2/2 (BI)		& 2	pers./column	& 2	pers./column		& 3							\\
\hline
\end{tabular}
\end{center}
\end{table}

With a loud oral 'start' signal pedestrians started walking in their respective directions simultaneously. Starting densities were specifically designed to allow a simultaneous start and we observed no significant differences between the movements of the first and last columns. Pedestrians walked through the measuring area and left it from their respective side. To avoid deviations in the results participants were asked to keep walking straight after leaving the measurement area and turn back toward their original starting area only after having completely walked the experimental section and the waiting area located at the opposite side (in other words no shortcuts were allowed).

\section{Inertial motion measurement using tablets}
In order to measure body movements of pedestrians and in particular their body rotation (sometimes referred as yawing), tablets equipped with a digital gyroscope were given to some of the participants (10 of them as indicated in Fig.3 and Fig.4). Nexus 7 (2013) tablets were chosen because of technical specifications and because of their relative large size and low weight, which allows to accurately follow body movements while having no (or little) influence on the walking behavior. A wearable bib with a zip pocket was used to fix the tablets on the chest of randomly selected participants. After wearing the bib it was tightly tightened (not too tight to still allow natural movements) with the tablets therefore measuring the motion of that portion of body between the chest and the belly. Participants chosen for measurement using tablets were of similar height and average build, making therefore comparison more easy and reliable.

\vspace*{4mm}
\centerline{\includegraphics[angle=90,trim=7cm 0cm 8cm 0cm,clip,width=11cm]{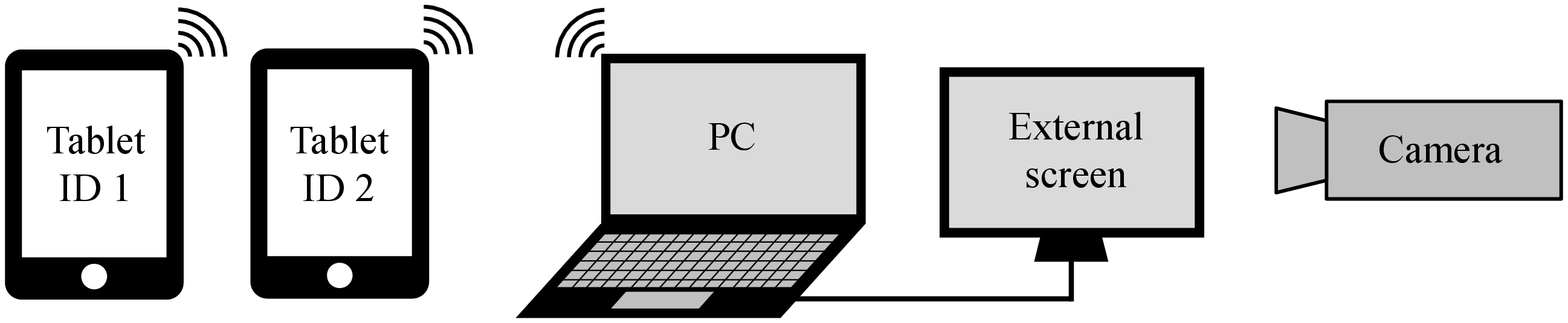}}
\begin{center}
\parbox{15.5cm}{\small\center{\bf Fig.5.} Data acquisition and time synchronization using tablets and PC. }
\end{center}
\vspace*{4mm}

An application reading the values generated from the gyroscope was installed on each tablet and sensing rate was set at 50 Hz. Because of the differences in clock time between each tablet a central storage was preferred. Inertial measurements from each tablet were sent over wireless network to a central PC where data were stored. This allowed to obtain the data from each tablet simultaneously, independently on the clock time differences between each tablet. \par
In addition, to allow synchronization with trajectory data obtained from video recording, an external monitor was connected to the central PC used to store inertial measurements. Camera's time and PC time were initially manually synchronized with an accuracy in the order of few seconds. To allow a more accurate synchronization the background color of the external screen was changed every 10 seconds (with the reference set at midnight), thus allowing to determine the PC time from the video recording with an accuracy of $\pm$ 1 frame. As a whole the synchronization error between the different tablets and the trajectories is of $\pm$ 1 frame, equivalent to about 0.03 s. A schematic representation of the technical system used to record, store and synchronize inertial movements measured from tablets is given in Fig.5.

\vspace*{4mm}
\centerline{\includegraphics[angle=0,trim=0.5cm 9cm 14cm 0.5cm,clip,width=12.5cm]{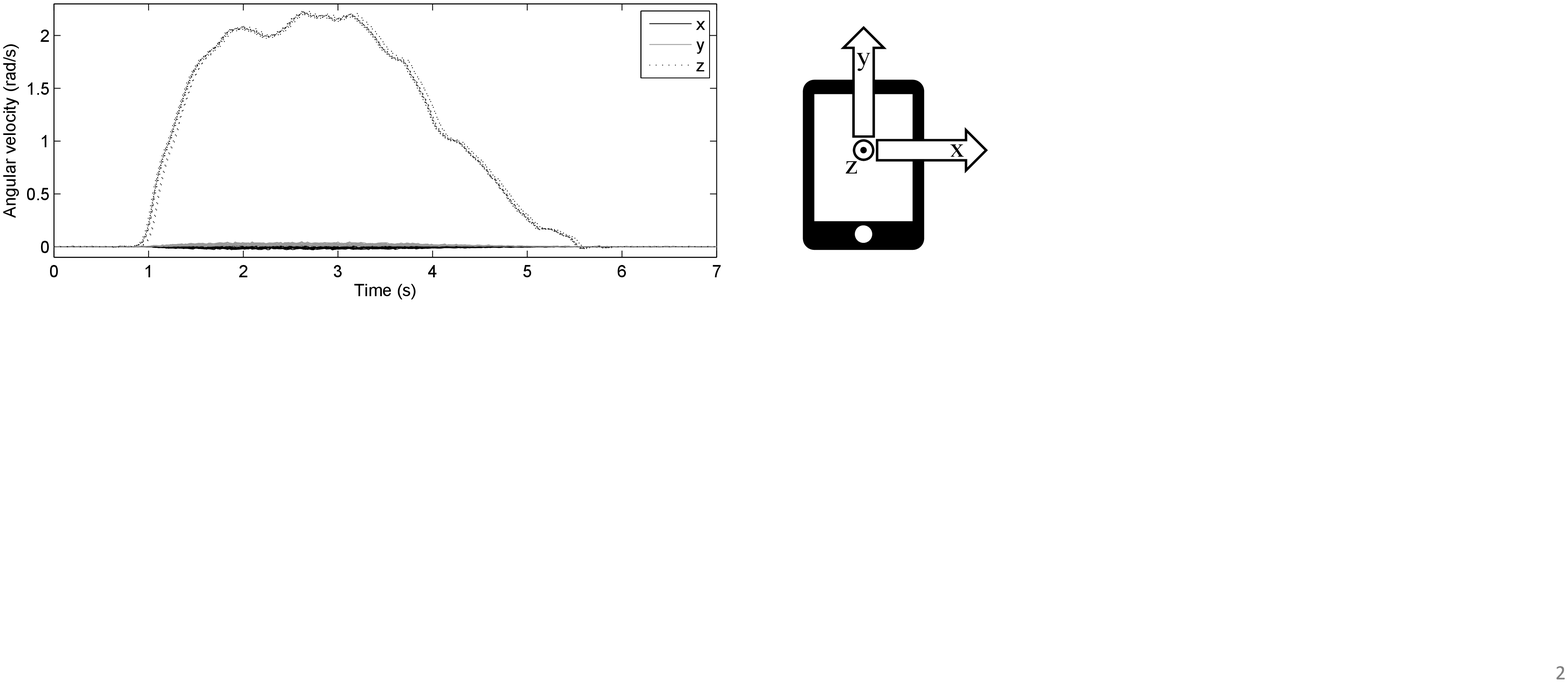}}
\begin{center}
\parbox{15.5cm}{\small\center{\bf Fig.6.} Angular velocity measurement validation (each line correspond to a different device). }
\end{center}
\vspace*{4mm}

Although the inertial sensor contained in tablets is used in a variety of applications (screen rotation, gaming, navigation,...), it is not, strictly speaking, a certified scientific instrument. We had therefore to check the accuracy and the precision in the measurement of angular velocity by using the internal gyroscope. For this scope we took the 10 tablets available, placed them in a tight box not allowing any type of independent movement and rotated them altogether by 360$^{\circ}$ using a rotating disk. As Fig.6 shows, a rotation on the tablet plane corresponds to a rotation in the $z$-axis relative to the coordinate of the gyroscope sensor. As the tablets are rotated, the angular velocity in the $z$-direction increases, before setting again to 0 once the rotation is stopped. The differences between the values recorded by each device are minimal (almost indistinguishable), giving an indication on the measurement precision. In addition, by integrating the signal obtained over the whole time period the angle measured can be obtained. In this case the recorded value was 360.33$\pm$1.53$^{\circ}$ suggesting that both precision and accuracy of the angular velocity measurement can be assessed as about 1\%.

\section{Results}
\subsection{Crossing time}
The analysis of the results can start by considering the average crossing time, i.e. the time required for a single pedestrian to cross the full length of the measurement area (10 m here). The results are shown in Tab.2 below and include all the repetitions indicated in Tab.1 and the full number of participants (50).

\begin{center}
Table 2: Crossing time and order parameter (see next section) during full bidirectional flow
\end{center}
\begin{table}[htbp]
\begin{center}
\begin{tabular}{|l|l|l|}
\hline 
Configuration name 					& Crossing time 					& Order parameter 		\\
\hline 
Unidirectional							& 6.643 $\pm$	0.258 s			& 1.000 $\pm$	0.000		\\
Unbalanced bidirectional 		& 7.471 $\pm$	0.803 s			& 0.980 $\pm$	0.040		\\
Balanced bidirectional 			& 7.772 $\pm$	0.910 s			& 0.958 $\pm$	0.051		\\
\hline
\end{tabular}
\end{center}
\end{table}

The unidirectional configuration has clearly the lowest crossing time, which had to be expected considering the fact that there are no head-on conflicts because all pedestrians move in the same direction. Additionally the unidirectional configuration shows the smallest dispersion, which means that the whole group moved in a fairly uniform way with small differences between each participant. The balanced bidirectional configuration was the least efficient one, both considering the high crossing time and the large dispersion.

\subsection{Analysis of pedestrians' trajectories}
In order to asses the efficiency of lanes in each configuration it is useful to introduce the order parameter, which allows to easily measure the efficiency or smoothness of bidirectional flows. But before continuing the presentation of the results it is important to define the 5 phases observed in the formation and dissolution of lanes in bidirectional flows. These phases can be summarized as follows:
\begin{enumerate}
\item Initial unidirectional free flow -- Pedestrians of both groups (major and minor flows) move toward each other's making up two unidirectional flows moving in opposite directions. Interactions between both groups, if present, are small and mostly of visual contact.
\item Lane formation -- Leading pedestrians of both groups first cross each other's. As both the major and the minor flow mix with each other's lanes are formed.
\item Full bidirectional flow -- Each layer of the measurement area can be regarded as an independent unidirectional flow. Pedestrians basically follow same-direction walkers and head-on collisions are drastically reduced.
\item Lane dissolution -- As the largest part of the group already left the central section, density is reduced and the increased surface allows more freedom in the movements thus contributing to the dissolution of ordered lanes.
\item Final unidirectional free flow -- When the last member of each group have passed each other's only unidirectional flows remain in each direction. In each half of the central section a unidirectional flow moving in opposite direction can be recognized. 
\end{enumerate}

The phases introduced here will be used to analyze more in detail the bidirectional flow.

\vspace*{4mm}
\centerline{\includegraphics[angle=0,trim=0cm 0cm 0cm 0cm,clip,width=6.5cm]{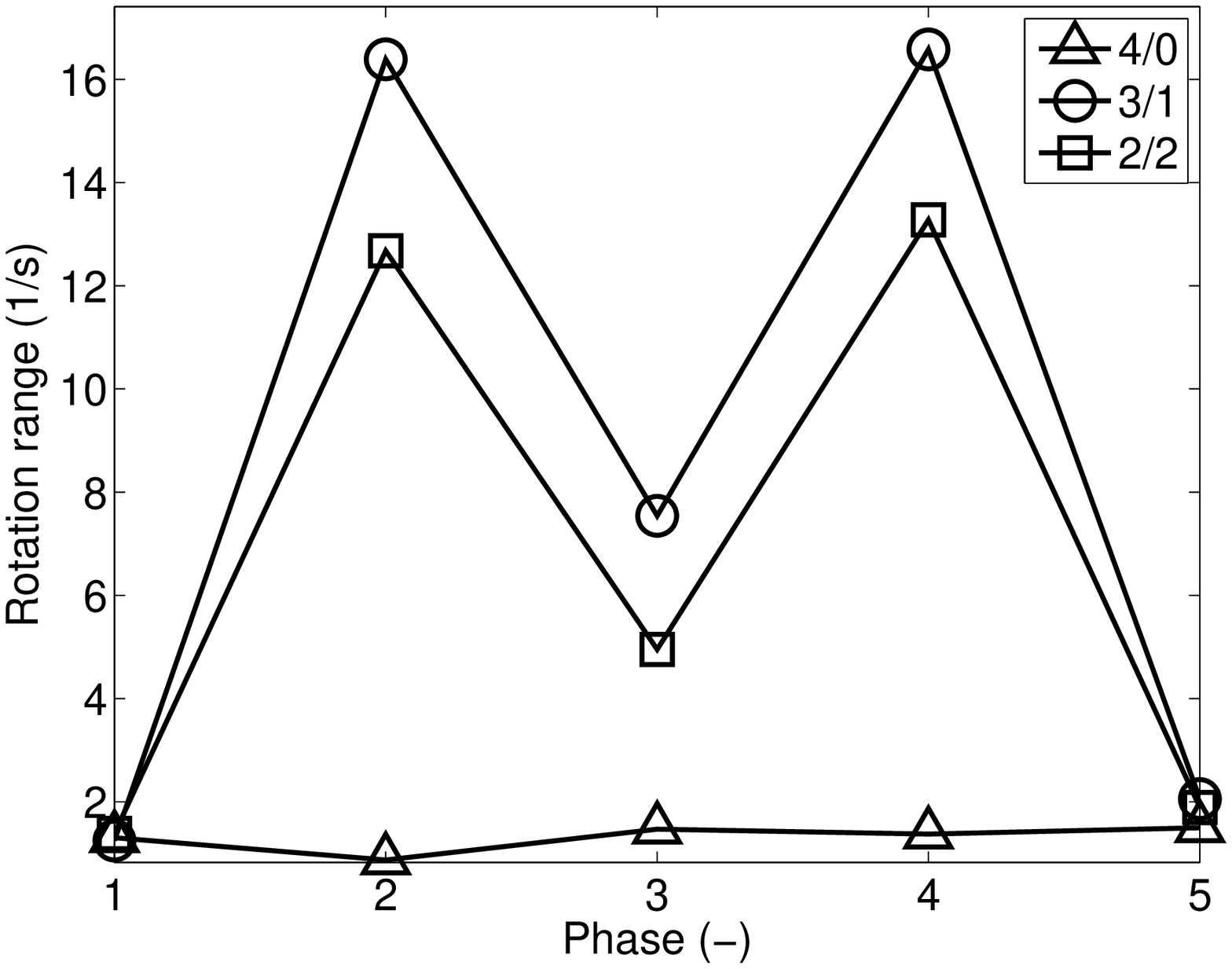}}
\begin{center}
\parbox{15.5cm}{\small\center{\bf Fig.7.} Flow rotation based on trajectories obtained from video processing. }
\end{center}
\vspace*{4mm}

Results for the order parameter in the 3 configurations are shown in Tab.2 and account for all the valid repetitions. The order parameter was computed using the trajectories obtained from the analysis of the video recordings by dividing the measurement area in a mesh having a cell width of 0.2 m (for details see \cite{Feliciani2016pre,Nowak2012}). Since order parameter is influenced by the left/right symmetry it is meaningful to consider only the fully developed bidirectional flow observed in phase 3. If a flow is perfectly organized, i.e. everyone moving in the same direction or in perfectly distinguished lanes, the order parameter becomes 1. For fully random motion, i.e. every pedestrian moving in a randomly chosen direction, it becomes 0. Every value included between the two extremes can be seen as a measure of efficiency for the lanes formed. \par
Because of its definition, the unidirectional flow has a constant order parameter equal to 1 during the whole length of the experiment and in all repetitions (uncertainty is 0). However, also the bidirectional cases show values very close to unity, meaning that very organized lanes were created. Differences between balanced and unbalanced cases are very small, making it difficult to distinguish both configurations in terms of smoothness of the lanes formed. \par
In addition it is possible to estimate the amount of rotation found in a given flow by employing a mathematical method (for details refer to \cite{Feliciani2016pre}). For each phase the measurement area is divided in a mesh of 0.2 m in size and the average velocity vector is computed in each cell. The rotation of the vector field can be computed by applying the curl operator on the whole measurement area. Since pedestrians velocities are only bi-dimensional (forward/backward and lateral motion) the resulting rotation vector will be perpendicular to the moving plane. The rotation range for each phase can be obtained by taking the difference between the maximum and minimum value of the vector field rotation in the measurement area. \par
Results for the rotation range are given in Fig.7. Phase 2 and 4 corresponding to lane formation and dissolution respectively show a very high rotation in both bidirectional cases, meaning that, in general, pedestrians tend to change their direction more frequently. During the initial and final phases low rotation range values are observed, meaning that a relatively smooth motion is observed (here all cases have only a unidirectional nature). In the full bidirectional phase (3) small rotation values are observed, suggesting that lane formation helps reducing the instability of the overall flow. During the whole length of the experiment unbalanced configuration shows larger rotation values compared to the balanced case. This is also true during the full bidirectional phase, although lanes of the unbalanced case were found being more organized. Considering both quantities together it may suggest that lanes formed in the unbalanced case were more organized in this experiment but less stable given the large values of rotation. 

\subsection{Body rotation}
Before introducing the experimental results for body rotation, it is important to explain the analysis method used to summarize data obtained from the gyroscope sensors used in this experiment. Fig.8 shows an example for the angular velocity measured on the body of a single pedestrian during the balanced bidirectional flow experiment. 

\vspace*{4mm}
\centerline{\includegraphics[angle=0,trim=4cm 14cm 6cm 1cm,clip,width=13.0cm]{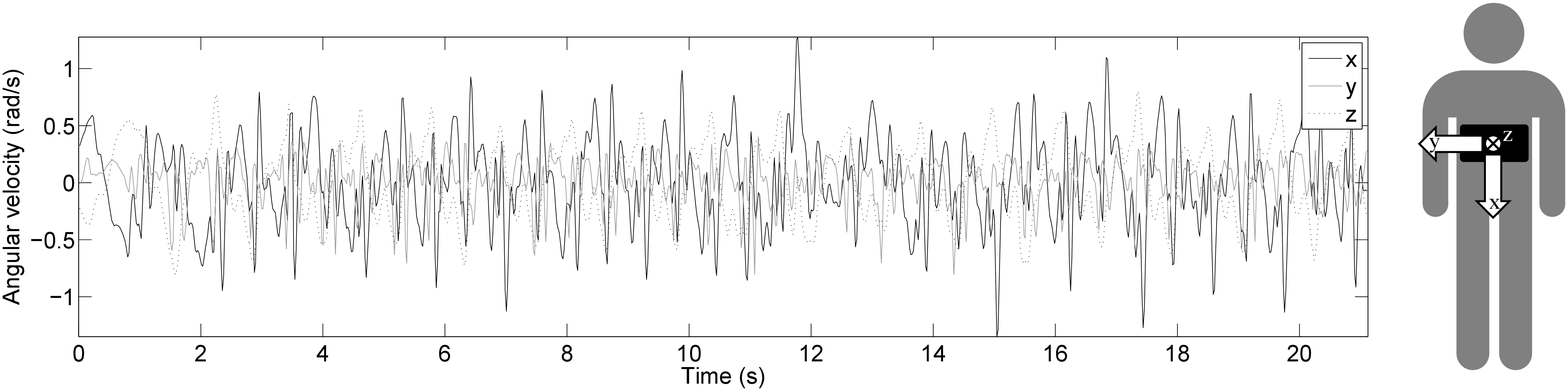}
						\hspace*{1mm}	
						\includegraphics[angle=0,height=3.5cm]{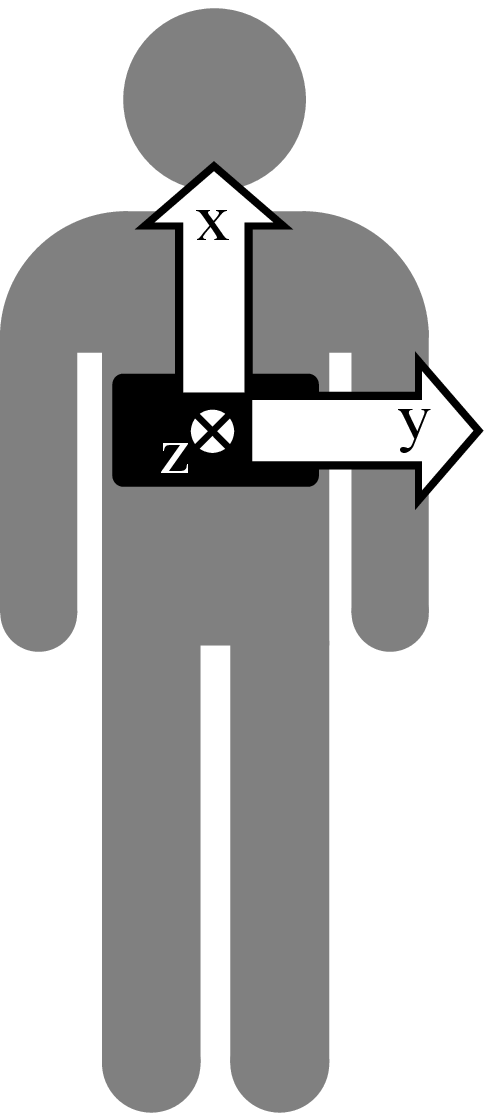}}
\begin{center}
\parbox{15.5cm}{\small\center{\bf Fig.8.} Example for the body angular velocity measured on a single pedestrian (balanced bidirectional flow). }
\end{center}
\vspace*{4mm}

As the figure shows it is difficult to distinguish a particular pattern in the graph of the recorded angular velocity. Clearly the alternating movement of legs has an influence in producing maximum and minimum with a characteristic frequency associated with the walking speed, but the graph created does not show a regular pattern, especially when the 3 components are considered altogether. To simplify the analysis and focus on the most relevant aspects found in bidirectional flows, we limited our considerations to the $x$-component, which stands for body rotation (or yawing). As Fig.8 shows, the $x$-component is the one having the largest amplitudes and considering the position of the tablets on the participants' body it is the one that can provide the most reliable information concerning walking attitude and upper body (shoulder) rotation. \par
But still, in order to compare the different flow configurations, analyzing the full angular velocity profile as a whole may be a difficult task. Since we are interested in a comparative/qualitative analysis we decided to use an average value defined as follows:

\begin{equation}
\overline{\omega}_x = \frac{\int_{t_0}^{t_1} | \omega_x |}{t_1-t_0} 
\end{equation}

where $\omega_x$ represents the angular velocity in $x$-direction while $t_0$ and $t_1$ are the integration limits (start- and end-time considered). In the case of the measured body rotation we were not able to distinguish between the different phases, because we could not determine individually the time at which pedestrians equipped with tablet entered and left the measurement area. Our analysis of the measured body rotation is therefore limited to the average value obtained by all participants equipped with tablets during the whole length of the experiment, defined as the time lasting from the first pedestrian entering the measurement area and the last leaving it. Directly measured body rotation was then compared with the flow rotation analysis performed by analyzing trajectories of the pedestrians. To compare both quantities an average and an integral values are used for the rotation range. In the first, the average value from Fig.7 is taken, while for the integral case each value is multiplied by the time length of the corresponding phase. Results are presented in Tab.3 below.

\begin{center}
Table 3: Directly measured body rotation and indirect measurement based on pedestrians' trajectories.
\end{center}
\begin{table}[htbp]
\begin{center}
\begin{tabular}{|l|l|l|l|}
\hline 
\multirow{2}{*}{Configuration} 	& \multirow{2}{*}{Yawing (gyroscope)} 	& \multicolumn{2}{c|}{Rotation range} 									\\
\cline{3-4}
																&																				& Average value									& Integral value				\\
\hline 
4/0 (UNI)												& 0.363 $\pm$ 0.046 rad/s								& 1.304 $\pm$ 0.360 s$^{-1}$		& 24.372 $\pm$ 7.647		\\
3/1 (BI)												& 0.430 $\pm$ 0.042 rad/s								& 8.763 $\pm$ 1.823 s$^{-1}$		& 176.785 $\pm$ 23.152	\\
2/2 (BI)												& 0.390 $\pm$ 0.056 rad/s								& 6.837 $\pm$ 2.223 s$^{-1}$		& 157.580 $\pm$ 48.343	\\
\hline
\end{tabular}
\end{center}
\end{table}

Although measuring principle (and units) are different, rotation measured on pedestrians' body is in agreement with the one estimated using trajectories. Both maximum and minimum values are found for the same configuration, with the unbalanced bidirectional configuration confirmed as rotating the most. While trend is similar, rotation based on trajectories seem more sensible on configuration changes, showing large differences where measured rotation only slightly changes. The relatively large uncertainty found in the balanced configuration may be related with the lowest number of repetitions (3 compared to 4 for the other cases).

\section{Conclusions}
In the present study we introduced a system to measure body rotation on pedestrians employing commercial tablets. The system was tested experimentally in a bidirectional flow environment by equipping 10 of the 50 participants with a tablet measuring body rotation by making use of the encapsulated digital gyroscope sensor. To check the accuracy and precision of the system a validation test was performed confirming that commercial devices can be used for scientific purposes with satisfying outcomes. \par
Results show that body rotation measured directly on pedestrians has a similar trend to the amount of rotation estimated using mathematical techniques based on pedestrians' trajectories. This suggests that quantities not directly measurable using image processing and/or tracking systems may be indirectly derived using alternative methods. For instance, body movement can be particularly difficult to be estimated based on direct information from video recording, because most of the methods rely on the recognition of pedestrians' head, which cannot provide a complete description of the movement being observed. However, some simple algorithm may be useful to get a first-instance estimation of the whole crowd like the rotation range presented here. \par
Finally, our analysis suggests that body rotation and/or oscillating movements may have a relationship with different emergent phenomena observed in pedestrian crowds and contribute to the stability of the structures formed. This qualitative assumption can be a motivation to develop more advanced simulation codes taking into consideration body shape and rotation in a more accurate manner. In more practical terms, it highlights the importance in providing an adequate guidance system in situations were critical flows are expected. In the particular case of bidirectional flow, by adequately separating streams moving in different directions and providing sufficient space before the interaction between each group, efficient lanes may form thus reducing congestion and related risks.

\section*{Acknowledgements}
This work was financially supported by JSPS KAKENHI Grant Number 25287026 and the Doctoral Student Special Incentives Program (SEUT RA) of the University of Tokyo. In addition the authors would like to thank the whole team of the Nishinari group for the assistance during the planning and the execution of the experiment reported here.

\bibliographystyle{IEEEtran}
\bibliography{PED2016_Feliciani}

\begin{thebibliography}{10}
\providecommand{\url}[1]{#1}
\csname url@samestyle\endcsname
\providecommand{\newblock}{\relax}
\providecommand{\bibinfo}[2]{#2}
\providecommand{\BIBentrySTDinterwordspacing}{\spaceskip=0pt\relax}
\providecommand{\BIBentryALTinterwordstretchfactor}{4}
\providecommand{\BIBentryALTinterwordspacing}{\spaceskip=\fontdimen2\font plus
\BIBentryALTinterwordstretchfactor\fontdimen3\font minus
  \fontdimen4\font\relax}
\providecommand{\BIBforeignlanguage}[2]{{%
\expandafter\ifx\csname l@#1\endcsname\relax
\typeout{** WARNING: IEEEtran.bst: No hyphenation pattern has been}%
\typeout{** loaded for the language `#1'. Using the pattern for}%
\typeout{** the default language instead.}%
\else
\language=\csname l@#1\endcsname
\fi
#2}}
\providecommand{\BIBdecl}{\relax}
\BIBdecl

\bibitem{Blue1999}
\BIBentryALTinterwordspacing
V.~Blue and J.~Adler, ``Cellular automata microsimulation of bidirectional
  pedestrian flows,'' \emph{Transportation Research Record: Journal of the
  Transportation Research Board}, no. 1678, pp. 135--141, 1999.  Available:
  \url{http://dx.doi.org/10.3141/1678-17}
\BIBentrySTDinterwordspacing

\bibitem{Helbing1995}
\BIBentryALTinterwordspacing
D.~Helbing and P.~Molnar, ``Social force model for pedestrian dynamics,''
  \emph{Physical review E}, vol.~51, no.~5, p. 4282, 1995.  Available:
  \url{http://dx.doi.org/10.1103/PhysRevE.51.4282}
\BIBentrySTDinterwordspacing

\bibitem{Weidmann1993}
\BIBentryALTinterwordspacing
U.~Weidmann, \emph{Transporttechnik der Fussg{\"a}nger: Transporttechnische
  Eigenschaften des Fussg{\"a}ngerverkehrs (Literaturauswertung)}.\hskip 1em
  plus 0.5em minus 0.4em\relax ETH, IVT, 1993.  Available:
  \url{http://dx.doi.org/10.3929/ethz-a-000687810}
\BIBentrySTDinterwordspacing

\bibitem{Lam2002}
\BIBentryALTinterwordspacing
W.~H.~K. Lam, J.~Y.~S. Lee, and C.~Y. Cheung, ``A study of the bi-directional
  pedestrian flow characteristics at hong kong signalized crosswalk
  facilities,'' \emph{Transportation}, vol.~29, no.~2, pp. 169--192, 2002.
  Available: \url{http://dx.doi.org/10.1023/A:1014226416702}
\BIBentrySTDinterwordspacing

\bibitem{Alhajyaseen2011}
\BIBentryALTinterwordspacing
W.~K.~M. Alhajyaseen, H.~Nakamura, and M.~Asano, ``Effects of bi-directional
  pedestrian flow characteristics upon the capacity of signalized crosswalks,''
  \emph{Procedia - Social and Behavioral Sciences}, vol.~16, pp. 526--535,
  2011.  Available: \url{http://dx.doi.org/10.1016/j.sbspro.2011.04.473}
\BIBentrySTDinterwordspacing

\bibitem{Kretz2008}
\BIBentryALTinterwordspacing
T.~Kretz, A.~Gr\"{u}nebohm, M.~Kaufman, F.~Mazur, and M.~Schreckenberg,
  ``Experimental study of pedestrian counterflow in a corridor,'' \emph{Journal
  of Statistical Mechanics}, 2006.  Available:
  \url{http://arxiv.org/abs/cond-mat/0609691}
\BIBentrySTDinterwordspacing

\bibitem{Kaufman2007}
M.~Kaufman, ``Lane formation in counterflow situations of pedestrian traffic,''
  Master's thesis, Universit\"{a}t Duisburg-Essen, 2007.

\bibitem{Nowak2012}
\BIBentryALTinterwordspacing
S.~Nowak and A.~Schadschneider, ``Quantitative analysis of pedestrian
  counterflow in a cellular automaton model,'' \emph{Physical Review E},
  vol.~85, no.~6, p. 066128, 2012.  Available:
  \url{http://dx.doi.org/10.1103/PhysRevE.85.066128}
\BIBentrySTDinterwordspacing

\bibitem{Feliciani2015}
\BIBentryALTinterwordspacing
C.~Feliciani and K.~Nishinari, ``Phenomenological description of deadlock
  formation in pedestrian bidirectional flow based on empirical observation,''
  \emph{Journal of Statistical Mechanics: Theory and Experiment}, vol. 2015,
  no.~10, p. P10003, 2015.  Available:
  \url{http://dx.doi.org/10.1088/1742-5468/2015/10/P10003}
\BIBentrySTDinterwordspacing

\bibitem{Zhang2012}
\BIBentryALTinterwordspacing
J.~Zhang, W.~Klingsch, A.~Schadschneider, and A.~Seyfried, ``Ordering in
  bidirectional pedestrian flows and its influence on the fundamental
  diagram,'' \emph{Journal of Statistical Mechanics: Theory and Experiment},
  vol. 2012, no.~02, p. P02002, 2012.  Available:
  \url{http://dx.doi.org/10.1088/1742-5468/2012/02/P02002}
\BIBentrySTDinterwordspacing

\bibitem{Feliciani2016}
\BIBentryALTinterwordspacing
{C. Feliciani and K. Nishinari}, ``An improved cellular automata model to
  simulate the behavior of high density crowd and validation by experimental
  data,'' \emph{Physica A: Statistical Mechanics and its Applications}, vol.
  451, pp. 135--148, 2016.  Available:
  \url{http://dx.doi.org/10.1016/j.physa.2016.01.057}
\BIBentrySTDinterwordspacing

\bibitem{Feliciani2016pre}
\BIBentryALTinterwordspacing
C.~Feliciani and K.~Nishinari, ``Empirical analysis of the lane formation
  process in bidirectional pedestrian flow,'' \emph{Physical Review E},
  vol.~94, p. 032304, 2016.  Available:
  \url{http://dx.doi.org/10.1103/PhysRevE.94.032304}
\BIBentrySTDinterwordspacing

\bibitem{Roggen2011}
\BIBentryALTinterwordspacing
D.~Roggen, M.~Wirz, G.~Tr{\"o}ster, and D.~Helbing, ``Recognition of crowd
  behavior from mobile sensors with pattern analysis and graph clustering
  methods,'' \emph{Physics and Society}, 2011.  Available:
  \url{http://arxiv.org/abs/1109.1664}
\BIBentrySTDinterwordspacing

\bibitem{Bujari2012}
\BIBentryALTinterwordspacing
A.~Bujari, B.~Licar, and C.~E. Palazzi, ``Movement pattern recognition through
  smartphone's accelerometer,'' in \emph{2012 IEEE Consumer Communications and
  Networking Conference (CCNC)}, 2012, pp. 502--506.  Available:
  \url{http://dx.doi.org/10.1109/CCNC.2012.6181029}
\BIBentrySTDinterwordspacing

\bibitem{Kjaergaard2012}
\BIBentryALTinterwordspacing
M.~B. Kj{\ae}rgaard, M.~Wirz, D.~Roggen, and G.~Tr{\"o}ster, ``Detecting
  pedestrian flocks by fusion of multi-modal sensors in mobile phones,'' in
  \emph{Proceedings of the 2012 ACM Conference on Ubiquitous Computing}, 2012,
  pp. 240--249.  Available: \url{http://dx.doi.org/10.1145/2370216.2370256}
\BIBentrySTDinterwordspacing

\bibitem{Boltes2013}
\BIBentryALTinterwordspacing
M.~Boltes and A.~Seyfried, ``Collecting pedestrian trajectories,''
  \emph{Neurocomputing}, vol. 100, pp. 127--133, 2013.  Available:
  \url{http://dx.doi.org/10.1016/j.neucom.2012.01.036}
\BIBentrySTDinterwordspacing

\end{thebibliography}

\end{document}